\documentclass[aps,english,prl,twocolumn,floatfix,superscriptaddress, showpacs]{revtex4}
\usepackage[T1]{fontenc}
\usepackage[latin9]{inputenc}
\usepackage{units}
\usepackage{graphicx}
\usepackage{esint}
\usepackage{babel}

\begin{document}

\title{Limits on inelastic dark matter from ZEPLIN-III}

\author{D.~Yu.~Akimov}
\affiliation{Institute for Theoretical and Experimental Physics, Moscow, Russia}
\author{H.~M.~Ara\'ujo}
\affiliation{Blackett Laboratory, Imperial College London, UK}
\author{E.~J.~Barnes}
\affiliation {School of Physics and Astronomy, University of Edinburgh, UK}
\author{V.~A.~Belov}
\affiliation{Institute for Theoretical and Experimental Physics, Moscow, Russia}
\author{A.~Bewick}
\affiliation{Blackett Laboratory, Imperial College London, UK}
\author{A.~A.~Burenkov}
\affiliation{Institute for Theoretical and Experimental Physics, Moscow, Russia}
\author{R.~Cashmore}
\affiliation{Brasenose College, University of Oxford, UK}
\author{V.~Chepel}
\affiliation{LIP--Coimbra \& Department of Physics of the University of Coimbra, Portugal}
\author{A.~Currie\footnote{Corresponding author, email: alastair.currie08@imperial.ac.uk}}
\affiliation{Blackett Laboratory, Imperial College London, UK}
\author{D.~Davidge}
\affiliation{Blackett Laboratory, Imperial College London, UK}
\author{J.~Dawson}
\affiliation{Blackett Laboratory, Imperial College London, UK}
\author{T.~Durkin}
\affiliation{Particle Physics Department, Rutherford Appleton Laboratory, Chilton, UK}
\author{B.~Edwards}
\affiliation{Particle Physics Department, Rutherford Appleton Laboratory, Chilton, UK}
\author{C.~Ghag}
\affiliation{School of Physics and Astronomy, University of Edinburgh, UK}
\author{A.~Hollingsworth}
\affiliation{School of Physics and Astronomy, University of Edinburgh, UK}
\author{M.~Horn}
\affiliation{Blackett Laboratory, Imperial College London, UK}
\author{A.~S.~Howard}
\affiliation{Blackett Laboratory, Imperial College London, UK}
\author{A.~J.~Hughes}
\affiliation{Particle Physics Department, Rutherford Appleton Laboratory, Chilton, UK}
\author{W.~G.~Jones}
\affiliation{Blackett Laboratory, Imperial College London, UK}
\author{G.~E.~Kalmus}
\affiliation{Particle Physics Department, Rutherford Appleton Laboratory, Chilton, UK}
\author{A.~S.~Kobyakin}
\affiliation{Institute for Theoretical and Experimental Physics, Moscow, Russia}
\author{A.~G.~Kovalenko}
\affiliation{Institute for Theoretical and Experimental Physics, Moscow, Russia}
\author{V.~N.~Lebedenko}
\affiliation{Blackett Laboratory, Imperial College London, UK}
\author{A.~Lindote}
\affiliation{LIP--Coimbra \& Department of Physics of the University of Coimbra, Portugal}
\affiliation{Particle Physics Department, Rutherford Appleton Laboratory, Chilton, UK}
\author{I.~Liubarsky}
\affiliation{Blackett Laboratory, Imperial College London, UK}
\author{M.~I.~Lopes}
\affiliation{LIP--Coimbra \& Department of Physics of the University of Coimbra, Portugal}
\author{R.~L\"{u}scher}
\affiliation{Particle Physics Department, Rutherford Appleton Laboratory, Chilton, UK}
\author{K.~Lyons}
\affiliation{Blackett Laboratory, Imperial College London, UK}
\author{P.~Majewski}
\affiliation{Particle Physics Department, Rutherford Appleton Laboratory, Chilton, UK}
\author{A.~StJ.~Murphy}
\affiliation {School of Physics and Astronomy, University of Edinburgh, UK}
\author{F.~Neves}
\affiliation{LIP--Coimbra \& Department of Physics of the University of Coimbra, Portugal}
\affiliation{Blackett Laboratory, Imperial College London, UK}
\author{S.~M.~Paling}
\affiliation{Particle Physics Department, Rutherford Appleton Laboratory, Chilton, UK}
\author{J.~Pinto da Cunha}
\affiliation{LIP--Coimbra \& Department of Physics of the University of Coimbra, Portugal}
\author{R.~Preece}
\affiliation{Particle Physics Department, Rutherford Appleton Laboratory, Chilton, UK}
\author{J.~J.~Quenby}
\affiliation{Blackett Laboratory, Imperial College London, UK}
\author{L.~Reichhart}
\affiliation {School of Physics and Astronomy, University of Edinburgh, UK}
\author{P.~R.~Scovell}
\affiliation {School of Physics and Astronomy, University of Edinburgh, UK}
\author{C.~Silva}
\affiliation{LIP--Coimbra \& Department of Physics of the University of Coimbra, Portugal}
\author{V.~N.~Solovov}
\affiliation{LIP--Coimbra \& Department of Physics of the University of Coimbra, Portugal}
\author{N.~J.~T.~Smith}
\affiliation{Particle Physics Department, Rutherford Appleton Laboratory, Chilton, UK}
\author{P.~F.~Smith}
\affiliation{Particle Physics Department, Rutherford Appleton Laboratory, Chilton, UK}
\author{V.~N.~Stekhanov}
\affiliation{Institute for Theoretical and Experimental Physics, Moscow, Russia}
\author{T.~J.~Sumner}
\affiliation{Blackett Laboratory, Imperial College London, UK}
\author{C.~Thorne}
\affiliation{Blackett Laboratory, Imperial College London, UK}
\author{L.~de Viveiros}
\affiliation{LIP--Coimbra \& Department of Physics of the University of Coimbra, Portugal}
\author{R.~J.~Walker}
\affiliation{Blackett Laboratory, Imperial College London, UK}

\collaboration{ZEPLIN-III}

\date{\today}

\begin{abstract}
We present limits on the WIMP-nucleon cross section for inelastic
dark matter derived from the 2008 run of ZEPLIN-III. Cuts, notably on scintillation
pulse shape and scintillation-to-ionisation ratio, give a net
exposure of $63\;\mathrm{kg \cdot days}$ in the range $20$--$80\;\mathrm{keV}$
nuclear recoil energy, in which 6 events are observed. Upper limits on signal rate are derived from the maximum empty patch in the data. Under standard halo assumptions a small region of parameter space consistent, at 99\% CL, with causing the $1.17\;\mathrm{ton \cdot year}$ DAMA modulation signal is allowed at 90\% CL: it is in the mass range $45$--$60\:\mathrm{GeV\;c^{-2}}$ with a minimum CL of 88\%, again derived from the maximum patch. This is the tightest constraint on that 
explanation of the DAMA result yet presented using a xenon target.
\end{abstract}


\maketitle
Dark matter in the form of weakly interacting massive particles (WIMPs)
which scatter predominantly into a higher-mass state has
been proposed \cite{Tucker-Smith2001} as an explanation of the annually
modulated event rate in DAMA/NaI and DAMA/LIBRA \cite{Bernabei2010} which is also
consistent with the upper limits on WIMP-nucleon elastic scattering rates from other
experiments \cite{Angle:2007uj,Lebedenko2009,Ahmed:2009zw}. In
such inelastic dark matter (iDM) models, scattering with energy
transfer $E_{\mathrm{R}}$ due to a WIMP of ground state mass $m_{\chi}$ and 
mass change $\delta$ requires a minimum relative speed \begin{equation}
v_{\mathrm{min}}=\frac{1}{\sqrt{2m\ensuremath{_{\mathrm{N}}}E_{\mathrm{R}}}}\left(\frac{m_{\mathrm{N}}E_{\mathrm{R}}}{\mu_{\mathrm{N}}}+\delta\right)\label{eq:vmin},\end{equation}
where $m_{\mathrm{N}}$ is the nucleus mass and $\mu_{\mathrm{N}}$ is the reduced
mass of the WIMP-nucleus system. A non-zero $\delta$ results in a
recoil spectrum that is zero at low energy and more sensitive, compared
with elastic scattering, to the upper tail of the WIMP velocity distribution.
WIMPs with velocity below $\left(2\delta/\mu_{\mathrm{N}}\right)^{0.5}$
will not scatter inelastically at all and so, for a given local escape velocity,
more $m_{\chi}$-$\delta$ parameter space is accessible
to heavier target nuclei. However, systematic uncertainty in the expected
relative rates in different targets due to nuclear form factors and WIMP velocity 
distributions grows with the difference in atomic mass \cite{Kuhlen2009}.
On balance, xenon is well suited to test iDM
models that would, by predicting a modulated rate of scattering against iodine nuclei, 
explain the DAMA observation.

ZEPLIN-III (described in detail in Refs.\ \cite{Akimov2007,Araujo2006})
is a liquid/gas detector designed to search for WIMPs scattering against
xenon nuclei in the 6.5~kg fiducial liquid volume. It is
built of low-radionuclide components, encased in hydrocarbon
and lead shielding, and operated in the Palmer Laboratory
at Boulby Mine beneath 2850~m water-equivalent rock overburden.
\begin{figure}[h]
\includegraphics[width=1\columnwidth]{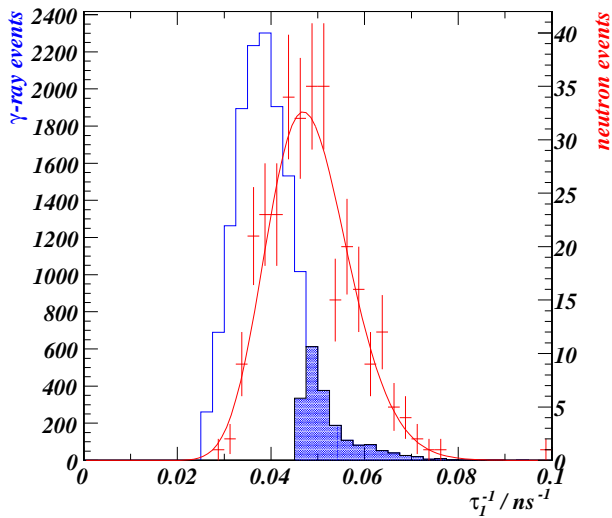}\caption{
The effect of the $\tau_{1}$ cut on the 20--25~keVee bin, over all $S2/S1$. The points and right scale 
correspond to AmBe data, fitted by a gamma distribution. The outlined histogram and left scale correspond to $^{137}$Cs data before the cut with a shaded region corresponding to the 13\% of electron recoil events which are not rejected by the cut.\label{fig:timing}}
\end{figure}
\begin{figure}[h]
\includegraphics[width=1\columnwidth]{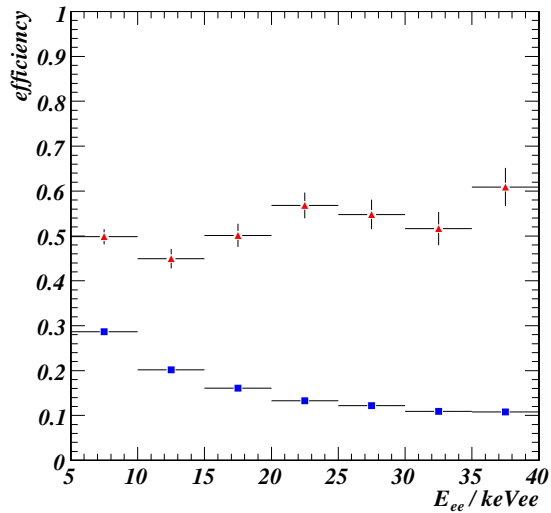}\caption{
The efficiency of the $\tau_{1}$ cut on calibration events over all $S2/S1$, for energy bins 5--40~keVee. Red triangles are 
AmBe data and blue squares are $^{137}$Cs data.\label{fig:tauEff}}
\end{figure}
Events are characterised by two light signals recorded by
an array of 31 photomultiplier tubes (PMTs). The summed scintillation
signal from the liquid is denoted by $S1$. A $3.9\;\mathrm{kV\, cm^{-1}}$ electric field in the 
liquid extracts ionisation charge from the interaction site, drifts it to the surface and 
forces emission into the gas layer above; there, an electroluminescence
signal, $S2$, is produced. As described in Ref.\ \cite{Lebedenko2009}, events with one $S1$ and one $S2$ signal were selected and cuts made, based on the pattern of light distribution, to remove multiple-scintillation, single-ionisation events.

An event's electron recoil equivalent energy, denoted by $E_{\mathrm{ee}}$ and measured in keVee, is derived from the pulse area of
the $S1$ signal, normalised to $122\;\mathrm{keV}$ photoabsorption
using a $^{57}$Co $\gamma$-ray source.
\begin{figure}[t]
\includegraphics[width=1\columnwidth]{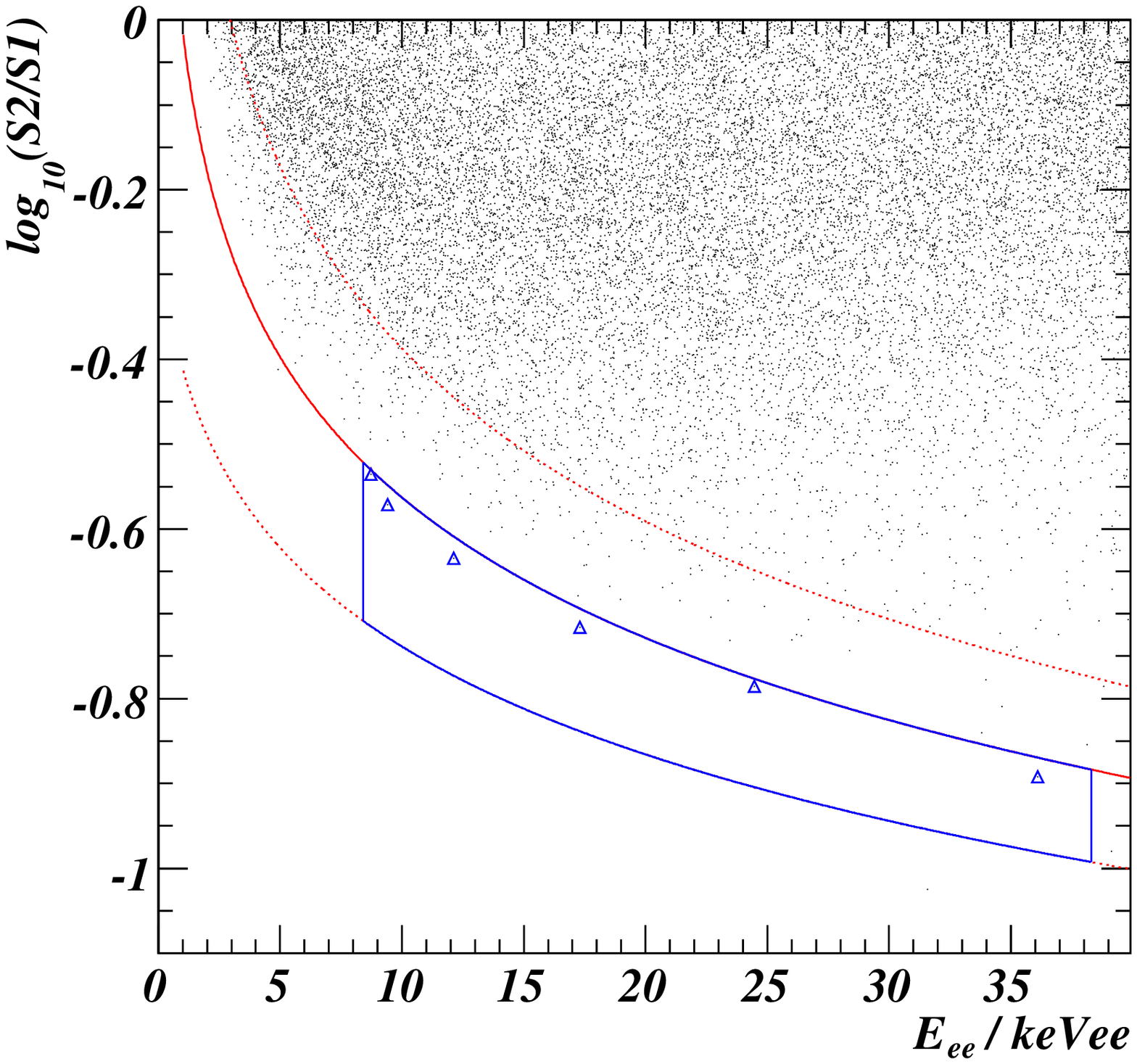}\caption{Search data passing all cuts except those on $E_{\mathrm{ee}}$ and $S2/S1$. Events passing all cuts are highlighted by triangles.
The solid and dashed lines show the mean plus and minus
two standard deviations for elastic AmBe calibration events, and the vertical lines indicate 20--80 keV nuclear recoil equivalent energy.\label{fig:search_data}}

\end{figure}
Discrimination between nuclear and electron recoil events is achieved primarily through the ratio
of scintillation and ionisation signals. Additional discrimination
has been achieved here using scintillation pulse shape. Recoiling electrons and nuclei produce different proportions 
of the singlet and triplet excited dimer states, which have lifetimes of 4 and 22 ns respectively \cite
{Hitachi:1983zz}. PMT traces are sampled at 2 ns intervals and so the mean arrival time of the $S1$ pulse area, 
denoted by $\tau_{1}$, is a useful discriminator. The timing of AmBe neutron calibration events within each 5-keVee 
bin from 5 to 40~keVee is well described by gamma 
distributions in $1/\tau_{1}$ \cite{Alner2005444}. Fitting a polynomial in $E_{\mathrm{ee}}$ to the medians of the gamma 
distributions produces a cut on $\tau_{1}$ with 50\% signal acceptance. Fig.\ \ref{fig:timing} shows the separation of the two recoil types and the effect of the cut in an 
example bin. The power of the timing cut to reduce electron recoil background increases with energy, as seen in Fig.
\ \ref{fig:tauEff}, mainly due to a narrowing in the $\tau_{1}$ distribution of electron recoil events.

AmBe calibration data were also used to obtain the $S2/S1$ distribution
of elastic nuclear recoil events which pass the timing cut, as a function of $E_{\mathrm{ee}}$. As in Ref.\ \cite
{Lebedenko2009}, the $\log_{10}\left(S2/S1\right)$ distribution was fitted by a 
Gaussian in each energy bin, and the energy dependence of the fitted means and standard deviations parametrised
by a power law to define a cut with 47.7\% signal acceptance. 
Charge recombination causes $S2$ and $S1$ to be microscopically anticorrelated at a given energy; in principle, therefore, the $S2/S1$ distribution has some dependence on the recoil energy spectrum. However, the low level of field-induced $S1$ suppression observed for nuclear recoils in xenon \cite{Manzur2010} suggests that the effect is relatively small. Here we have assumed, as xenon experiments historically have, that the $S2/S1$ distribution at fixed $S1$ for neutron calibration events is an adequate approximation to that for signal events. After 
efficiencies from dead time, pulse-finding, event reconstruction and the cuts on $S2/S1$ and $\tau_{1}$, the net
 exposure for signal events is $63\:\mathrm{kg\cdot days}$, with 5\% uncertainty due to neutron calibration statistics.

Nuclear recoil-equivalent energy, $E_{\mathrm{R}}$, is determined as in Ref.\ \cite{Lebedenko2009} from $E_{\mathrm{ee}}$ via a conversion factor:
\begin{equation}
E_{\mathrm{R}}=\frac{S_{\mathrm{e}}}{L_{\mathrm{eff}}S_{\mathrm{n}}}E_{\mathrm{ee}},\end{equation}
where $S_{\mathrm{e}}$ and $S_{\mathrm{n}}$ are the field-induced
suppression factors for the light yield of electron and nuclear
recoils and $L_{\mathrm{eff}}$ is the
zero-field light yield of nuclear recoils relative to that of electron recoils. An energy range 
of 20--80 keV nuclear recoil energy (8.4--38.3 keVee) was chosen 
to include the majority of events predicted by the quenched,
inelastic WIMP-iodine scattering interpretation of the DAMA modulation \cite{Chang2009}.

Fig.\ \ref{fig:search_data} shows the six search events which passed all cuts. The combined efficiency of the cuts 
on $S2/S1$ and $\tau_{1}$ for search data (6 in $1.3\times10^{5}$) is no higher than for electron-recoil 
calibration data (7 in $8.5\times10^{4}$), suggesting that the surviving search events may well constitute the tail 
of the electron-recoil background population. Without the timing cut the box would have contained 27 events.

\begin{figure}[t]
\includegraphics[width=1\columnwidth]{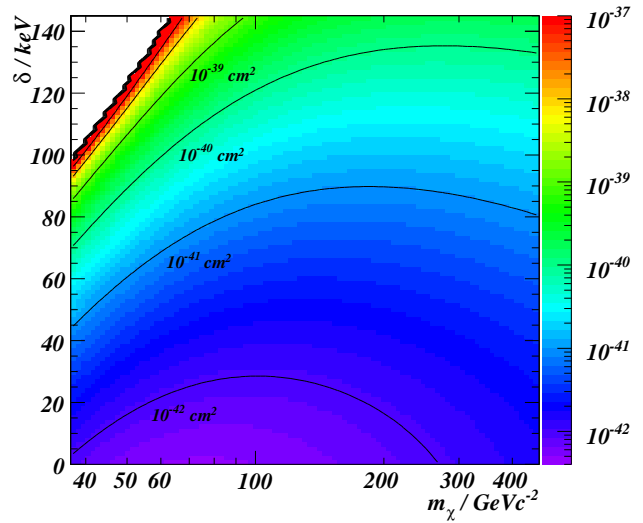}\caption{(color online) 90\% 
limit on $\sigma_{n}/\mathrm{cm^{2}}$
as a funciton of mass and splitting. The upper left region predicts no inelastic scattering during the run.
\label{fig:sigmaLimit}}

\end{figure}

For WIMPs which couple equally to protons and neutrons, the differential
rate for spin-independent WIMP-nucleus scattering in a target
of total mass $M_{\mathrm{T}}$ is given by:
\begin{equation}
\frac{dR}{dE_{\mathrm{R}}}\left(E_{\mathrm{R}},t\right)=\frac{M_{\mathrm{T}}\rho_{\chi}\sigma_{n}}{2m_{\chi}\mu_{\mathrm{n}}^{2}}A^{2}F^{2}(q)\int_{\mathrm{v_{min}}}^{\infty}d^{3}v\frac{f(\vec{v},t)}{v},\end{equation}
where $\rho_{\chi}$ is the local WIMP density, $A$ is the atomic
number of the target nucleus, $\sigma_{n}$ is the WIMP-nucleon cross
section, $\mu_{\mathrm{n}}$ is the WIMP-nucleon reduced mass 
and $f\left(\vec{v},t\right)$ is the WIMP velocity distribution
in the target frame. A Helm form factor was used: \begin{equation}
F(q)=\frac{3j_{1}\left(qr_{n}\right)}{qr_{n}}\exp\left(-(qs)^{2}/2\right),\end{equation}
 for momentum transfer $q$, where the effective nuclear radius is
taken to be $r_{n}=\sqrt{1.44A^{\frac{2}{3}}-5}\;\mathrm{fm}$, the
skin depth $s=1\;\mathrm{fm}$ and $j_{1}$ is a spherical Bessel
function.

Recoil energy spectra were calculated under a standard halo model: 
$\rho_{\chi}=0.3\;\mathrm{GeV}\,\mathrm{c}^{-2}\,\mathrm{cm^{-3}}$,
a Maxwellian velocity distribution with $v_{0}=220\;\mathrm{m\, s^{-1}}$
truncated at escape velocity $v_{\mathrm{esc}}$ in the
galactic frame, and an Earth velocity parametrised as in Ref.\ \cite{Gelmini2001}.
The underlying spectrum for given $m_{\chi},\delta$ and $\sigma_{n}$
was modified by the energy resolution and efficiency of ZEPLIN-III
and then averaged over the 83-day run to produce a signal model. The
energy resolution, dominated by Poisson statistics of photoelectron production and 
the variance of the single-photoelectron response, is 
$\sigma/E_{\mathrm{R}} = 1.5\left(E_{\mathrm{R}}/\mathrm{keV}\right)^{-0.5}$.

The maximum patch statistic \cite{Henderson2008} was used to
derive single-sided upper limits on the rate of signal events in
the 20--80 keV range. No background estimate is used; consequently,
the null hypothesis cannot be ruled out by this method. Events were mapped onto a plane of uniform signal density by integrating the signal spectrum in $E_{\mathrm{R}}$ and the fitted profile in $S2/S1$. For models in the previously un-excluded region of iDM parameter space, the largest empty rectangle in the re-mapped search box has a fractional acceptance of 0.73--0.75; this implies a 90\% CL limit of 5.4--5.1 expected signal events in the box. The resultant
limits on $\sigma_{n}$ for $v_{\mathrm{esc}}=550\;\mathrm{km\,s^{-1}}$ are plotted in 
Fig.\ \ref{fig:sigmaLimit}.

\begin{figure*}[t!]
\includegraphics[width=0.33\textwidth]{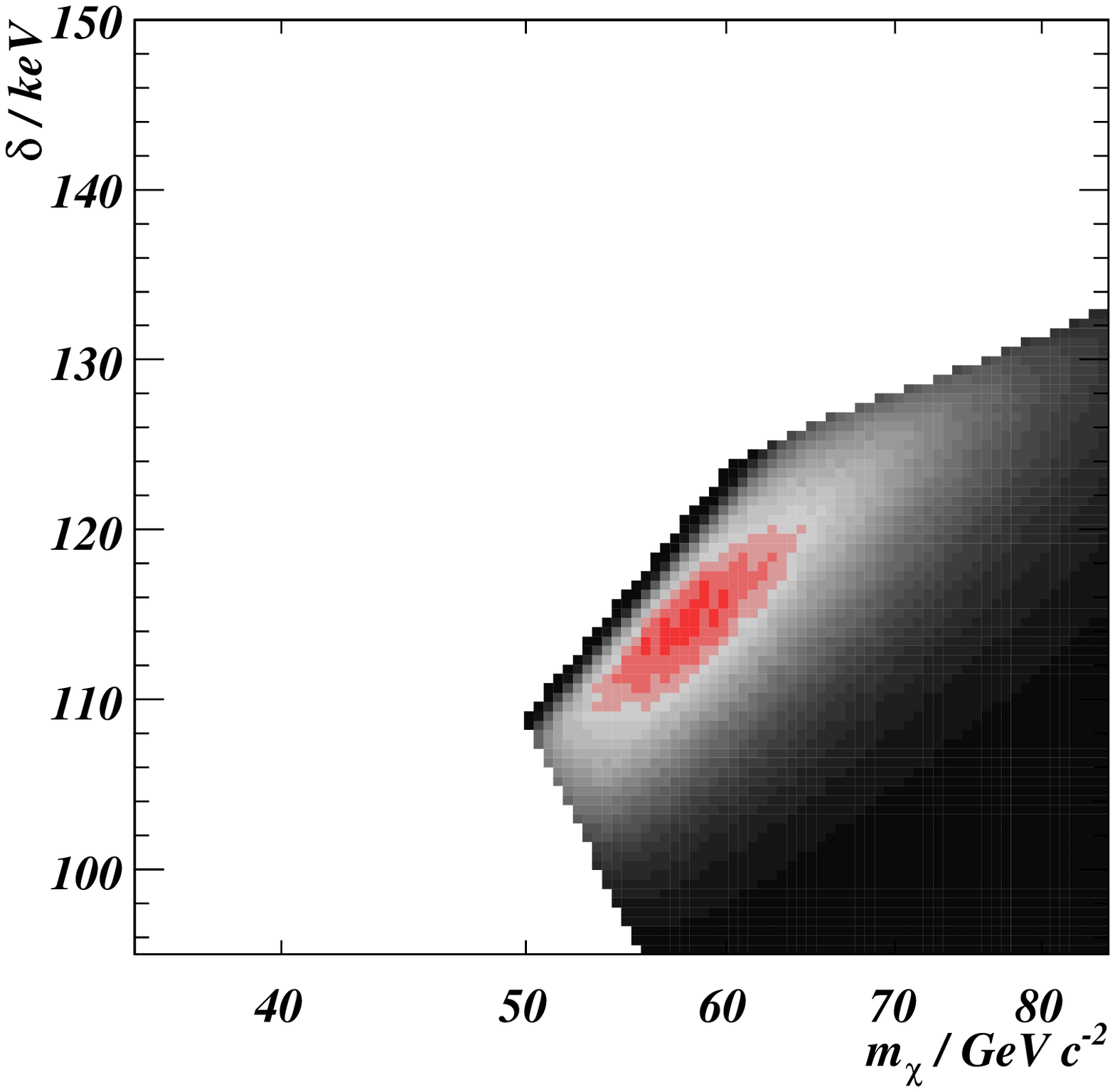}\includegraphics[width=0.33\textwidth]{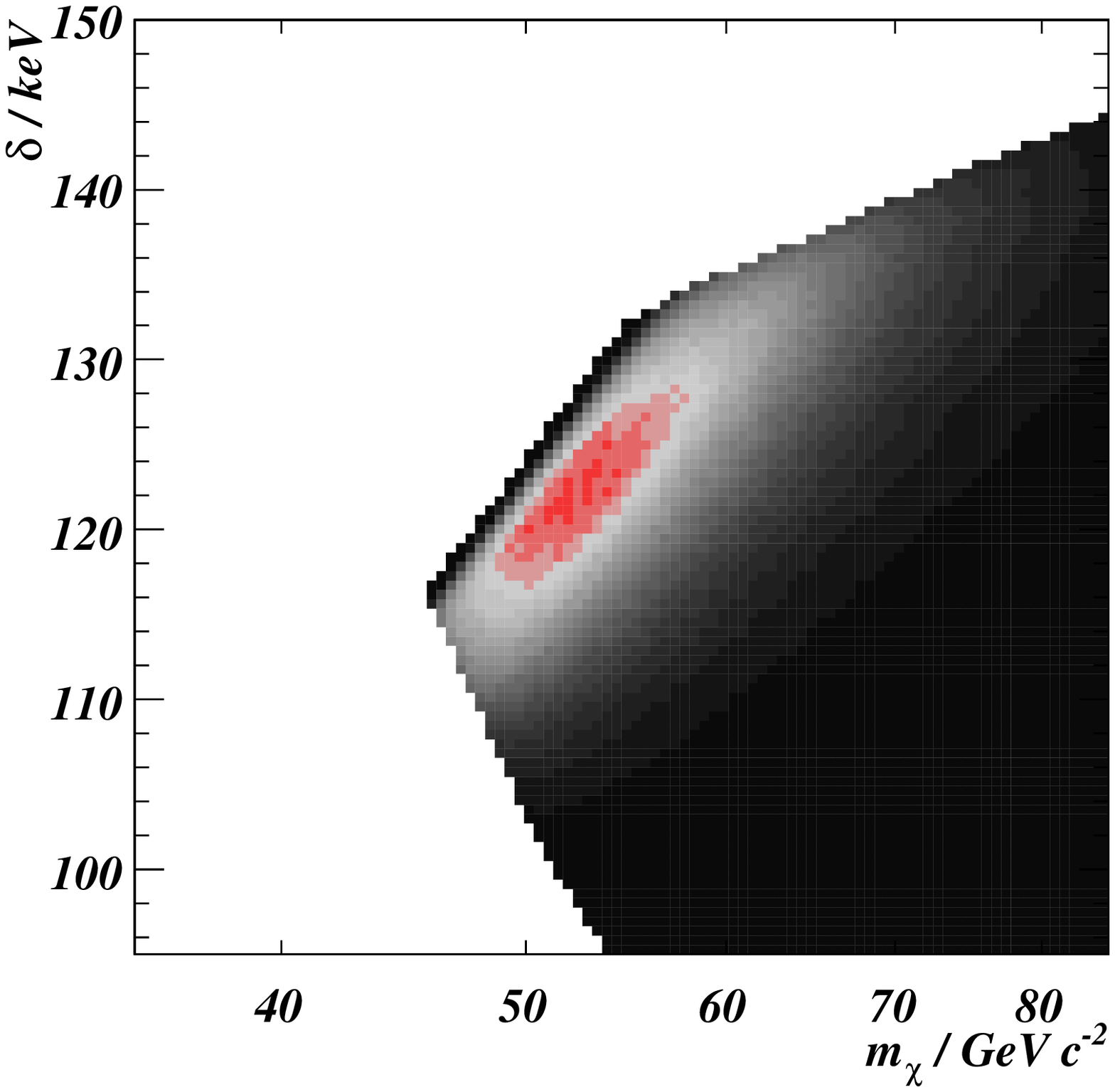}\includegraphics[width=0.33\textwidth]{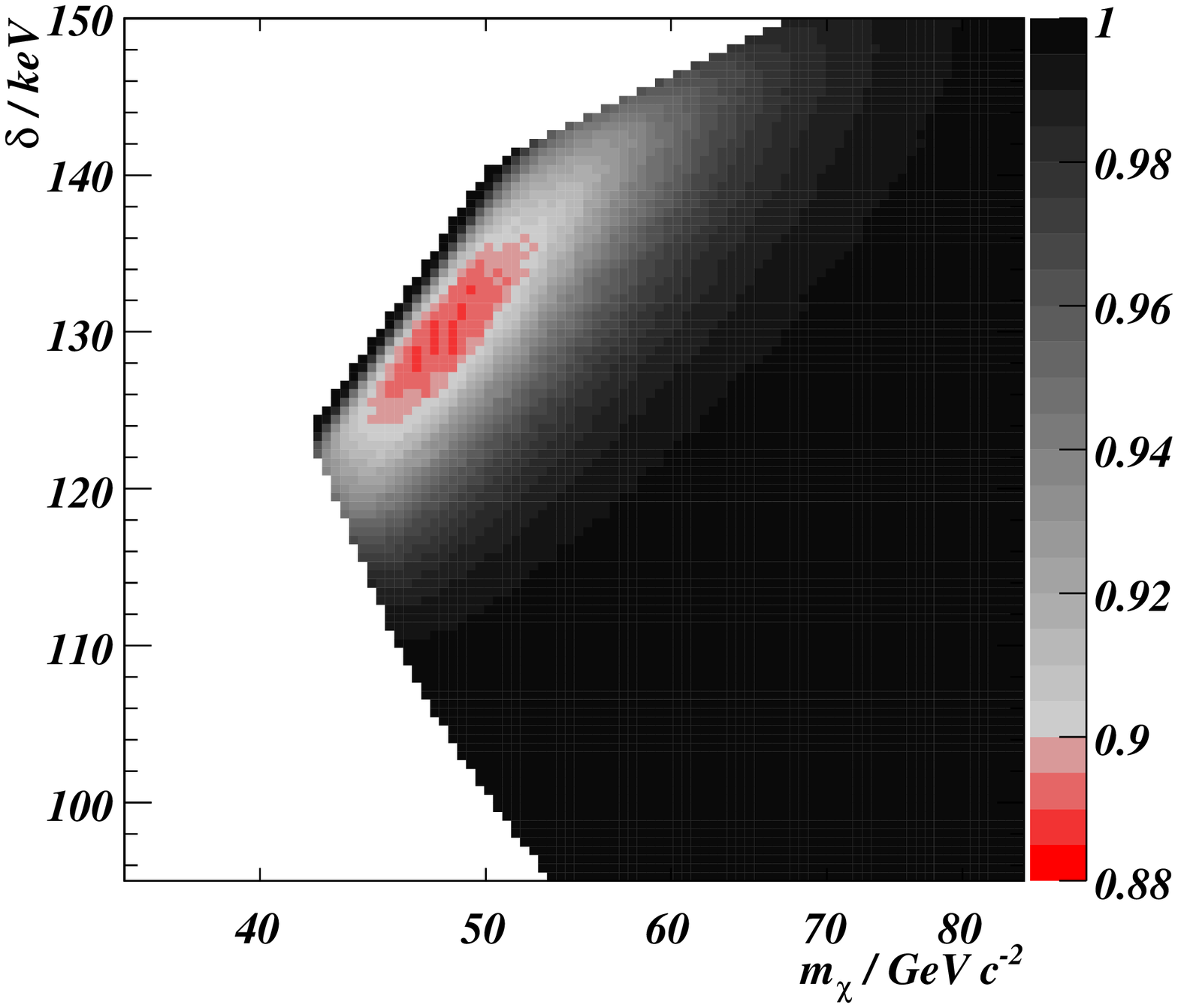}\caption{In $m_{\chi}$--$\delta$ space, the confidence level at which ZEPLIN-III excludes the lowest value of $\sigma_{n}$ consistent, at 99\% CL, with causing the DAMA modulation. Three values of $v_{\mathrm{esc}}$ are shown: (from left) 500, 550 and 600 $\mathrm{km\; s^{-1}}$.\label{fig:allowedRegions}}

\end{figure*}

Signal modulation spectra for the combined DAMA experiments were constructed
with resolution as described in Refs.\ \cite{Bernabei1999,Bernabei:2007c}
and parametrised in Ref.\ \cite{Schmidt-Hoberg2009}. An iodine quenching factor of 0.08 \cite{Gerbier1999, Tovey1998} was used; the exclusion results are relatively insensitive to channelling effects \cite{Bernabei2008a} which are, conservatively, omitted. The parameters $m_{\chi}$,
$\delta$ and $\sigma_{n}$ were fitted, by minimizing $\chi^{2}$,
to the observed modulation amplitude in 0.5-keVee bins from 2--10
keVee and a single 10--20 keVee bin, following Ref.\ \cite{Schmidt-Hoberg2009}. A 90\% confidence interval for the local 
escape velocity from Ref.\ \cite{Smith:2006ym} is 498--608 km s$^{\mathrm{-1}}$ and the cross section excluded by 
ZEPLIN-III depends on the true $v_{\mathrm{esc}}$. Non-Maxwellian velocity distributions would cause a similar 
systematic effect. Fig.\ \ref{fig:allowedRegions} shows the ZEPLIN-III constraints on parameter space consistent with causing the DAMA modulation for three values of $v_{\mathrm{esc}}$. DAMA-explaining cross sections are excluded at the 88\% confidence level. Fluctuations of $\pm1\cdot\sigma$ in the cut efficiencies derived from neutron calibration would change this minimum CL within the range 87--90\%.

In summary, a search of $63\:\mathrm{kg\cdot days}$ net exposure with a xenon target yielded 6 candidate events in the range 20--80 keV nuclear recoil equivalent energy. They were consistent, both in number and scintillation-to-ionisation ratio, with belonging to the tail of an electron recoil background population. Single-sided upper limits were set on the WIMP-nucleon cross section, constraining the DAMA-explaining region of iDM parameter space: for a standard halo model there remains a 90\% CL allowed region for WIMP masses in the range $45$--$60\:\mathrm{GeV\;c^{-2}}$, with minimum CL 88\%. This is more stringent than limits from other xenon and germanium experiments \cite{Cline:2009xd,Angle2009,Ahmed:2009zw} and supports previous exclusions \cite{Schmidt-Hoberg2009} based on CRESST-II data. In particular, a target element of similar mass to iodine reduces systematic uncertainty due to the WIMP velocity distribution.

The UK groups acknowledge the support of the Science \& Technology
Facilities Council for the ZEPLIN-III project and for
maintenance and operation of the underground Palmer laboratory which
is hosted by Cleveland Potash Ltd at Boulby Mine. The project would not be
possible without the cooperation of the management and staff of CPL. The authors 
thank Neal Weiner for helpful discussions.
We acknowledge support from a Joint International Project
award, held at ITEP and ICL, from the Russian Foundation of Basic
Research (08-02-91851 KO~a) and the Royal Society. LIP-Coimbra acknowledges 
financial support from Funda\c{c}\~{a}o para Ci\^{e}ncia e Tecnologia 
through the project grants CERN/FP/83501/2008 and CERN/FP/109320/2009, as well 
as postdoctoral grants SFRH/BPD/27054/2006 and SFRH/BPD/47320/2008. This 
work was supported in part by SC Rosatom; by Russian Grant SS-1329.2008.2 and 
by the Russian Ministry of Education and Science contract 02.740.11.0239. The 
University of Edinburgh is a charitable body, registered in Scotland, with 
the registration number SC005336.

\bibliographystyle{apsrev}
\bibliography{iDMpaper}

\end{document}